\documentclass[conference]{IEEEtran}
\IEEEoverridecommandlockouts
\usepackage{cite}
\usepackage{amsmath,amssymb,amsfonts}
\usepackage{algorithmic}
\usepackage{graphicx}
\usepackage{textcomp}
\usepackage{xcolor}
\usepackage{url}
\usepackage{tcolorbox}
\def\BibTeX{{\rm B\kern-.05em{\sc i\kern-.025em b}\kern-.08em
    T\kern-.1667em\lower.7ex\hbox{E}\kern-.125emX}}
\begin{document}

\title{Automated Fault Detection in 5G Core Networks Using Large Language Models
}

\author{\IEEEauthorblockN{Parsa Hatami\textsuperscript{*}}
\IEEEauthorblockA{\textit{Department of Electrical Engineering} \\
\textit{Sharif University of Technology}\\
Tehran, Iran \\
parsa.hatami81@sharif.edu}
\and
\IEEEauthorblockN{Ahmadreza Majlesara\textsuperscript{*}}
\IEEEauthorblockA{\textit{Department of Electrical Engineering} \\
\textit{Sharif University of Technology}\\
Tehran, Iran \\
ahmad.ara@sharif.edu}
\and
\IEEEauthorblockN{Ali Majlesi}
\IEEEauthorblockA{\textit{Department of Electrical Engineering} \\
\textit{Sharif University of Technology}\\
Tehran, Iran \\
ali.majlesi@sharif.edu}
\and
\IEEEauthorblockN{Babak Hossein Khalaj}
\IEEEauthorblockA{\textit{Department of Electrical Engineering} \\
\textit{Sharif University of Technology}\\
Tehran, Iran \\
khalaj@sharif.edu}
\vspace{0.5em}
\textsuperscript{*}These authors contributed equally to this work.

}

\maketitle

\begin{abstract}
With the rapid growth of data volume in modern telecommunication networks and the continuous expansion of their scale, maintaining high reliability has become a critical requirement. These networks support a wide range of applications and services, including highly sensitive and mission-critical ones, which demand rapid and accurate detection and resolution of network errors. Traditional fault-diagnosis methods are no longer efficient for such complex environments.\cite{b1}
In this study, we leverage Large Language Models (LLMs) to automate network fault detection and classification. Various types of network errors were intentionally injected into a Kubernetes-based test network, and data were collected under both healthy and faulty conditions. The dataset includes logs from different network components (pods), along with complementary data such as system descriptions, events, Round Trip Time (RTT) tests, and pod status information. The dataset covers common fault types such as pod failure, pod kill, network delay, network loss, and disk I/O failures.
We fine-tuned the GPT-4.1 nano model via its API on this dataset, resulting in a significant improvement in fault-detection accuracy compared to the base model. These findings highlight the potential of LLM-based approaches for achieving closed-loop, and operator-free fault management, which can enhance network reliability and reduce downtime-related operational costs for service providers.
\end{abstract}

\begin{IEEEkeywords}
fault detection, 5G networks, root cause analysis, Large Language Models (LLMs), Kubernetes
\end{IEEEkeywords}

\section{Introduction}
Maintaining the continuous operation of modern communication networks—especially advanced infrastructures such as 5G systems—has become both crucial and increasingly challenging. Traditional network monitoring approaches rely heavily on manual inspection of logs, alerts, and performance metrics. Such processes are time-consuming, error-prone, and lack scalability as networks grow in size, complexity, and heterogeneity\cite{b2}. As telecommunication systems continue to evolve, integrating software-defined architectures, virtualization, and containerized services, the rapid identification and mitigation of faults have become even more difficult to achieve.

Recent advancements in artificial intelligence (AI), and particularly in Large Language Models (LLMs), offer a new paradigm for intelligent and automated network management \cite{b3} \cite{b4} \cite{b6}. These models possess the ability to understand natural language, interpret unstructured data such as logs, and even reason about causal relations in complex systems. This capability introduces a significant opportunity to transform traditional network operations—from reactive troubleshooting to proactive, self-healing mechanisms. While earlier studies explored AI-driven monitoring through statistical and deep learning models\cite{b2}\cite{b5}, LLMs now enable contextual reasoning across multimodal data sources, thereby bridging the gap between human expertise and automated fault diagnosis.

With the global transition from 4G to 5G and beyond, the scale and diversity of telecommunication infrastructures have expanded dramatically. Network environments now integrate cloud-native platforms, such as Kubernetes, where numerous microservices interact dynamically. Configuring, managing, and troubleshooting these environments demand both domain-specific and software-engineering expertise. Leveraging LLMs to automate such complex management tasks has therefore gained significant attention in recent research and industrial discussions, including works such as LLM for 5G: Network Management\cite{b7}. By combining domain knowledge with the reasoning capabilities of LLMs, it becomes feasible to realize autonomous, closed-loop network operations capable of detecting, classifying, and resolving errors in near real time.
\\
A review of previous research reveals that no prior work has explored a generalized framework for assessing the overall health of communication networks using Large Language Models (LLMs) to identify potential faults across different layers and components. Most existing studies focus on specific layers—such as the physical or data-link layer—and rely on structured numerical data rather than unstructured system information. Techniques such as anomaly detection and time-series forecasting have been applied successfully\cite{b1}\cite{b2} in these limited contexts to detect deviations in predefined performance indicators.

In parallel, several works have investigated log analysis using traditional machine-learning or rule-based approaches\cite{b4} \cite{b5} \cite{b6}. However, these methods typically operate on pre-parsed and highly filtered logs within controlled experimental conditions, requiring significant preprocessing efforts. As a result, their applicability is restricted to detecting a narrow range of explicitly defined errors that appear in standardized log formats. While such techniques can achieve high accuracy in constrained settings, they are not scalable or adaptable to heterogeneous, dynamic network environments.


Recent domain-specific efforts have targeted 5G fault detection and RCA with hybrid ML–LLM pipelines. In the core network, \emph{5G Core Fault Detection and Root Cause Analysis using Machine Learning and Generative AI} combines supervised anomaly detectors with LLM-based summarization to improve diagnosis quality on control/user-plane events, demonstrating that generative models can convert heterogeneous operational traces into operator-ready incident narratives~\cite{b8}. Complementary trends extend beyond the core toward cross-domain or cross-vendor settings: \emph{An LLM-based Cross-Domain Fault Localization in Carrier Networks} shows that aligning prompts with topology/context priors improves generalization across environments, while \emph{RCA Copilot} operationalizes LLMs as a copilot over multi-source telemetry (logs, metrics, traces), yielding actionable recommendations rather than raw detections~\cite{b9,b10}. These systems substantiate the utility of LLMs for interpretability and triage, yet most still depend on structured features, curated parsers, or offline post-processing pipelines, which limits real-time applicability in highly dynamic, cloud-native 5G cores.

A parallel line of work studies LLMs (and tool-augmented LLM agents) for automated RCA and decision support. \emph{RCAgent} integrates tool calls and knowledge retrieval with conversational reasoning to correlate events and surface root causes in complex cloud stacks~\cite{b11}. In telecom-adjacent 5G contexts, LLM-driven RCA has also been explored for RAN anomalies using graph/transformer models to represent dependencies, and for verticals such as power-grid 5G where domain knowledge must be fused with network telemetry~\cite{b12,b13}. \emph{Reasoning Language Models for Root Cause Analysis in 5G Wireless Networks} further argues for explicit chain-of-thought/plan-and-act prompting to improve causal attribution under noisy observability~\cite{b14}. Across these works, three gaps remain: (i) reliance on pre-parsed or schema-constrained inputs rather than raw, heterogeneous logs and events; (ii) limited coupling with live orchestrators (e.g., Kubernetes) for closed-loop response; and (iii) scarcity of fine-tuned datasets built from realistic fault injections in operational 5G cores.


In this study, our objective is to design a comprehensive and flexible pipeline capable of processing heterogeneous network inputs—such as raw logs, events, and status reports—without extensive preprocessing or manual parsing. The proposed LLM-based approach acts as an intelligent conversational assistant that enables network operators, even without deep technical expertise, to monitor and diagnose network conditions using simple natural-language queries. Furthermore, the adaptability of LLMs allows new data types or fault categories to be seamlessly integrated into the model, dynamically expanding its operational scope. Finally, this model can be deployed in a closed-loop, real-time monitoring framework, continuously observing the network, identifying anomalies, and generating detailed diagnostic reports whenever faults occur. Such an approach has the potential to revolutionize network management by combining automation, interpretability, and reliability within a single unified system\cite{b3}\cite{b7}.

\section{METHODOLOGY}
\begin{figure}[h]
\centerline{\includegraphics[width=3.3in]{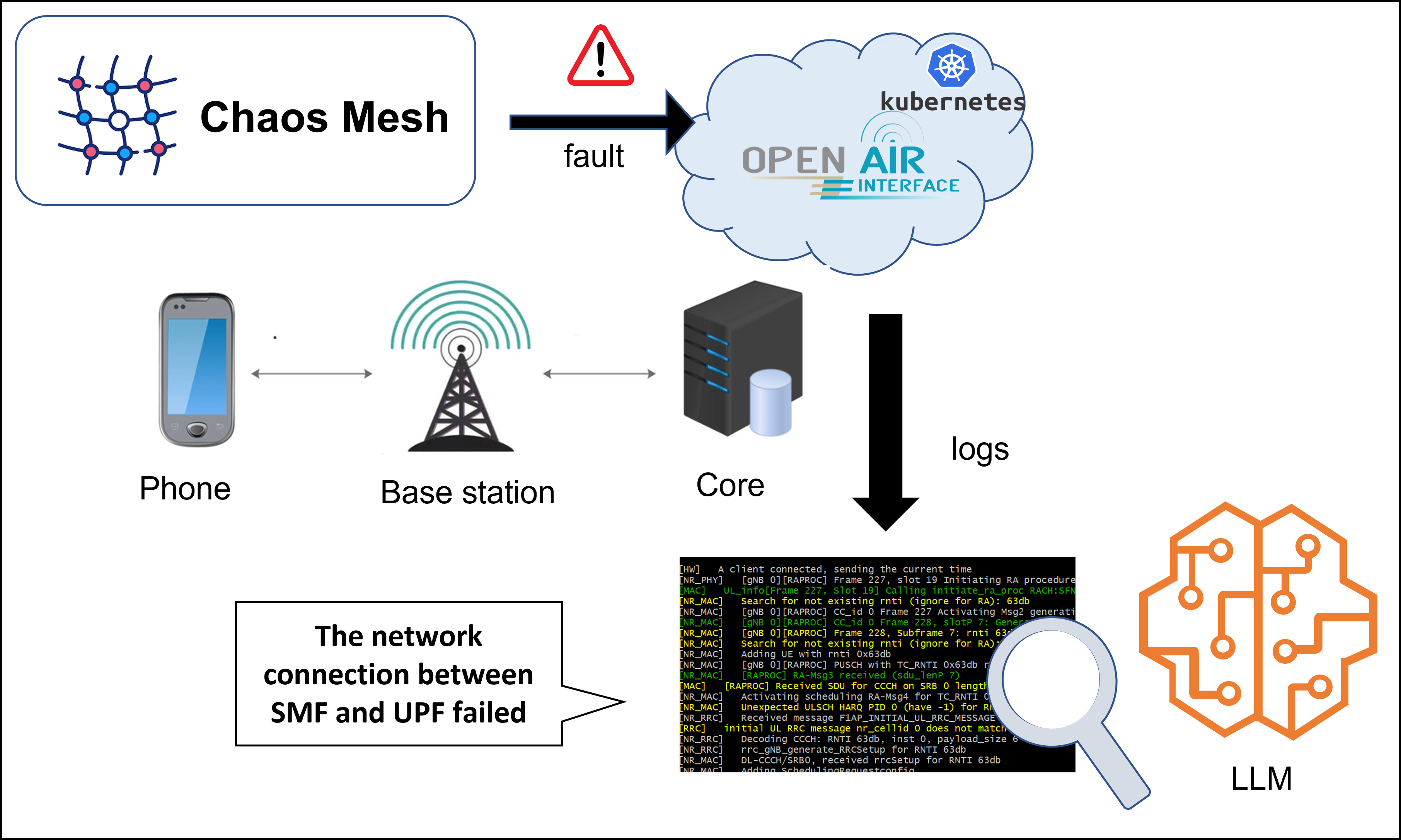}}
\caption{Dataset generation for training an LLM on fault detection using Chaos Mesh.}
\label{fig:entire_pipeline}
\end{figure}

An LLM is trained on logs from 5G mobile networks to detect the root cause of failures. For training, faults are generated using Chaos Mesh in an OpenAirInterface 5G network deployed on a Kubernetes cluster. The entire training pipeline is shown in Fig.~\ref{fig:entire_pipeline}.

\subsection{OpenAirInterface 5G Core}

To evaluate the proposed framework under realistic network conditions, we developed an experimental 5G testbed based on the OpenAirInterface (OAI) 5G Core, an open-source and standards-compliant implementation of the 3GPP Release 15 specification\cite{b15}. 

Our core network includes the Access and Mobility Management Function (AMF), the Session Management Function (SMF), the User Plane Function (UPF). A dedicated database module maintains subscriber and session information. The radio access layer of the testbed consists of an OAI-based gNodeB (gNB) connected to multiple User Equipments (UEs), forming a fully functional RAN-Core integration.

In our implementation, each network component was deployed as a containerized pod within a Kubernetes cluster, allowing for isolated management, automated scaling, and continuous log collection. These logs, along with performance metrics from the gNB and UEs, served as the primary data source for our LLM-based monitoring and fault-detection pipeline.

\begin{figure}[h]
\centerline{\includegraphics[width=3.3in]{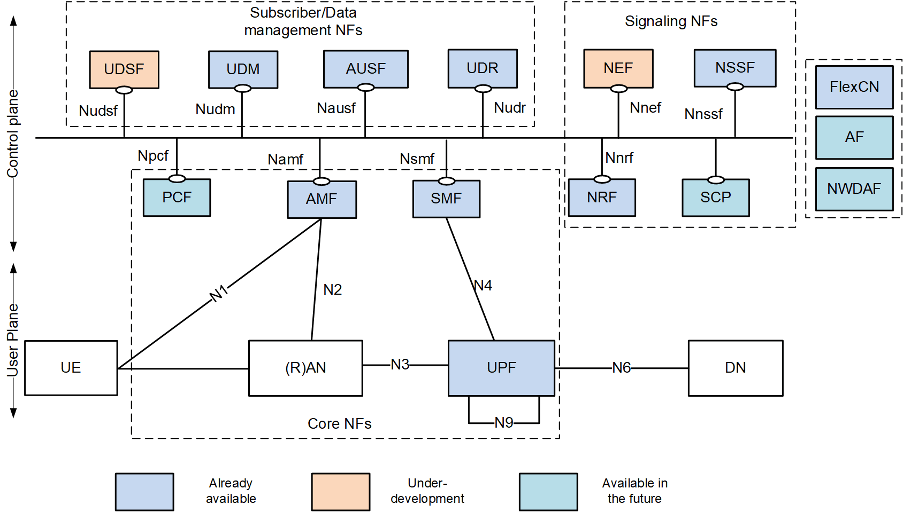}}
\caption{OpenAirInterface 5G Core Architecture Diagram (adapted from the official OAI documentation \cite{b15}).}
\label{fig:OAI}
\end{figure}

\subsection{Chaos Mesh}

To generate dataset for the proposed LLM-based monitoring framework, controlled fault scenarios were introduced into the experimental testbed using Chaos Mesh, a powerful open-source chaos engineering platform designed for Kubernetes environments\cite{b16}. Chaos Mesh enables the simulation of various failure conditions at the pod, network, and system levels in a safe and repeatable manner.
In this study, five representative categories of faults were selected to emulate common failure conditions in cloud-native 5G systems:

1) Pod Failure:
Simulates unexpected termination or crash of a containerized network function, such as AMF or SMF. This fault type tests the resilience of service recovery mechanisms and the system’s ability to restart or reassign affected pods automatically.

2) Pod Kill:
Represents a forced termination of a specific pod process at runtime. Unlike pod failure, this type mimics operator-triggered or abrupt shutdown scenarios, providing insights into how quickly the orchestration layer (Kubernetes) detects and replaces the terminated component.

3) Network Delay:
Introduces artificial latency into communication between selected pods or components (e.g., between the gNB and the UPF). This allows evaluation of the system’s performance under delayed signaling and degraded transport conditions, reflecting scenarios such as congested backhaul links.

4) Network Loss:
Simulates packet drops within network interfaces to analyze the tolerance of the 5G core control and data planes to unreliable transmission. This experiment helps study how message retransmissions and session stability are affected under partial network loss.

5) I/O Injection:
Injects read/write latency or temporary disk unavailability into containerized components, replicating storage-related anomalies such as database I/O bottlenecks or temporary filesystem access failures. This fault type examines the robustness of stateful services like the core database module.

Each fault type was applied to different components of the OAI 5G Core for predefined time intervals, while comprehensive telemetry data—including system logs, event traces, pod status updates, and round-trip time (RTT) measurements—were collected in parallel. This process yielded a rich dataset containing both normal and abnormal operational states, which was subsequently used for fine-tuning the LLM and validating its fault detection and reasoning capabilities.

Through this controlled and systematic approach, Chaos Mesh provided a reliable means to evaluate the network’s fault response and to train the LLM model under realistic and diverse failure conditions.

\subsection{Data Gathering}

To build a training and evaluation dataset for our LLM-based fault diagnosis system, we developed an automated data collection pipeline that interacts directly with the Kubernetes-based 5G core deployment. The pipeline remotely controls the testbed, orchestrates experiments, injects faults, and captures multi-source telemetry before and after each fault scenario in a structured and repeatable manner. The multi-source telemetry collection and aggregation design is inspired by cloud-native observability frameworks such as Prometheus and Kubernetes logging best practices.\cite{b17}

The data collection process is executed from a controller machine that connects to the cluster over SSH. Through this interface, the pipeline programmatically issues operational commands to query the state of the system, retrieve runtime information, and persist relevant outputs. The workflow consists of four main stages:

\begin{enumerate}
    \item \textbf{Network Reset and Initialization:} At the start of each experiment, the testbed is brought into a known healthy baseline state. The script tears down and reinstalls the mobile core network (AMF, SMF, UPF, database, gNB) and user equipment (UE) instances using predefined deployment descriptors. After redeployment, the pipeline automatically waits until all critical components are reported as ready by Kubernetes. This ensures that each experiment begins from a consistent configuration and that observed degradations are attributable to injected faults rather than residual transient behavior from previous runs.

    \item \textbf{Health Monitoring and Status Capture:} The script periodically inspects the lifecycle state of all pods (e.g., \texttt{Running}, \texttt{CrashLoopBackOff}, etc.) to detect any non-running or partially degraded components. Pod status summaries are stored to disk at each experiment step. In addition, a ``describe'' snapshot is collected for every pod. This \texttt{kubectl describe pod} output includes recent events, restart counts, container-level conditions, resource usage signals, and warning messages generated by Kubernetes. Capturing these descriptions is critical for post hoc reasoning, because they often contain early indicators of faults (for example, repeated restarts, image pull errors, failed liveness probes) that may not yet appear in high-level KPIs.

    \item \textbf{Runtime Telemetry and Connectivity Measurements:} For each core network function and UE pod, the pipeline retrieves recent logs directly from the workload containers. These logs include authentication traces, session setup messages, mobility management events, forwarding plane messages, and error messages from services such as AMF, SMF, and UPF. In addition to log data, the pipeline actively measures connectivity quality between the core and each UE by executing an RTT (round-trip time) test. RTT results are collected for all UE instances both before and after the fault is injected. This allows us to correlate control-plane or user-plane anomalies with observable service impact at the UE side.

    \item \textbf{Event and System-Level Reporting:} Alongside pod-local telemetry, the pipeline also records the global Kubernetes event history in the target namespace. The event trace captures cluster-level reactions to faults (e.g., pod eviction, rescheduling attempts, container kills, liveness probe failures, node-level resource pressure). By including these events in the dataset, we provide the model with causal context: not only what failed, but also how the orchestration layer responded.
\end{enumerate}

All collected artifacts from a single experimental run---pod logs, pod descriptions, per-pod status, UE RTT measurements (before/after fault), and namespace-wide event history---are stored for each experiment.

This automated approach has three important properties. First, it guarantees repeatability: each dataset sample corresponds to a known network state and a known injected fault condition. Second, it captures heterogeneous data sources (raw logs, status descriptions, cluster events, RTT measurements) without requiring heavy manual preprocessing or hand-written parsing rules. Third, it produces operator-facing, human-readable traces that can be directly consumed by an LLM. In other words, rather than relying solely on numerical metrics or narrowly structured telemetry, the pipeline aggregates semantically rich, text-based observations that reflect how real engineers would inspect and debug a live 5G core deployment.

These consolidated experiment snapshots form the supervision data used for fine-tuning the LLM. Each snapshot is paired with the ground-truth fault label associated with that experiment (e.g., \textit{network delay}, \textit{pod kill}, \textit{I/O degradation}), enabling supervised training for fault classification and explanatory diagnosis.

\subsection{Log Filtering}

Log files are typically very large. For instance, within four minutes, the gNodeB generates over 2,500 log lines (amounting to more than 60,000 tokens), costing about \$0.001 when processed using \texttt{gpt-4o-mini}. Therefore, the logs are first filtered to retain only the important lines.

\begin{figure}[h]
\centerline{\includegraphics[width=3.3in]{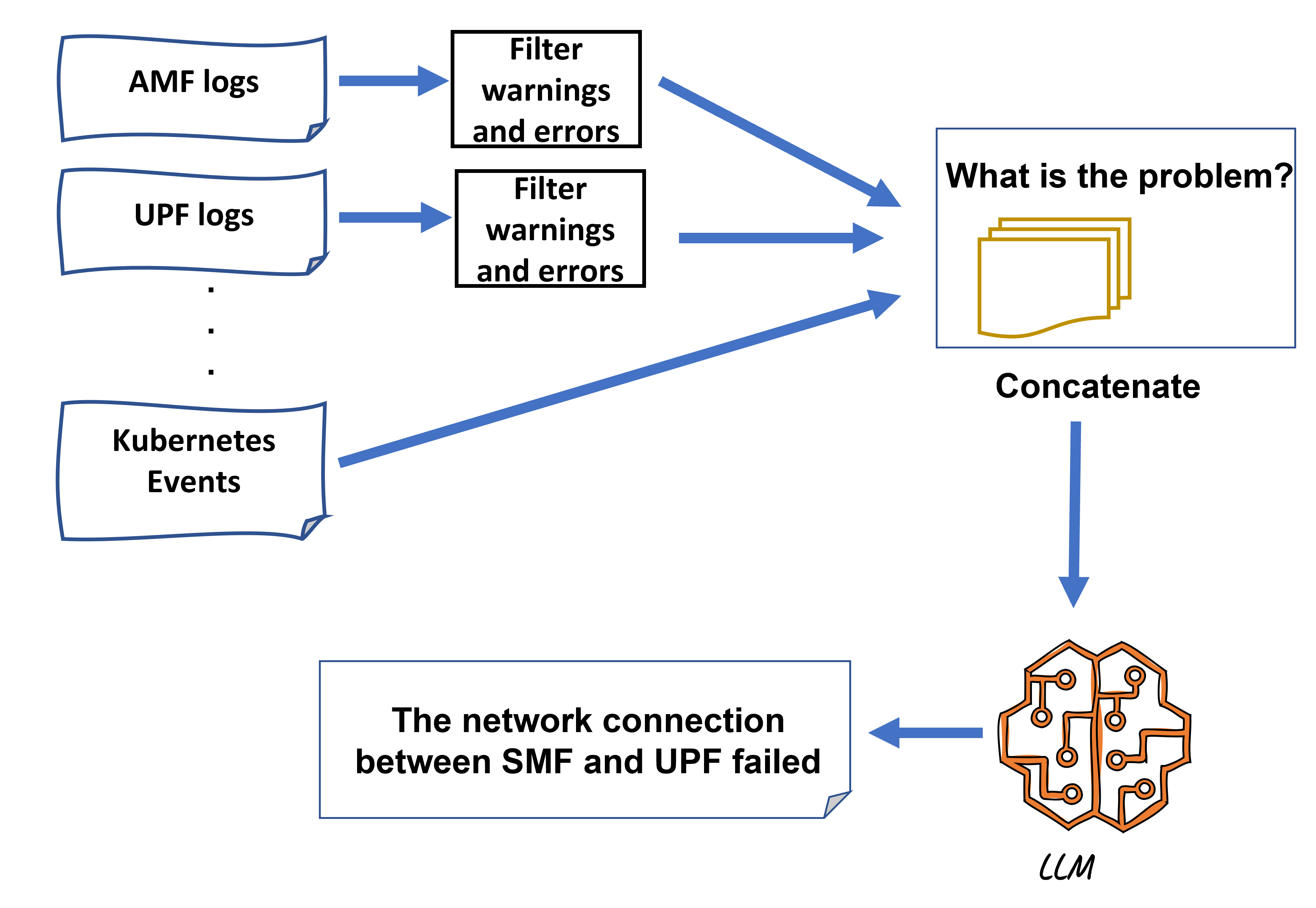}}
\caption{Fault Detection Pipeline}
\label{fig:fault_detection_pipeline}
\end{figure}

\begin{figure}[h]
\centerline{\includegraphics[width=3.3in]{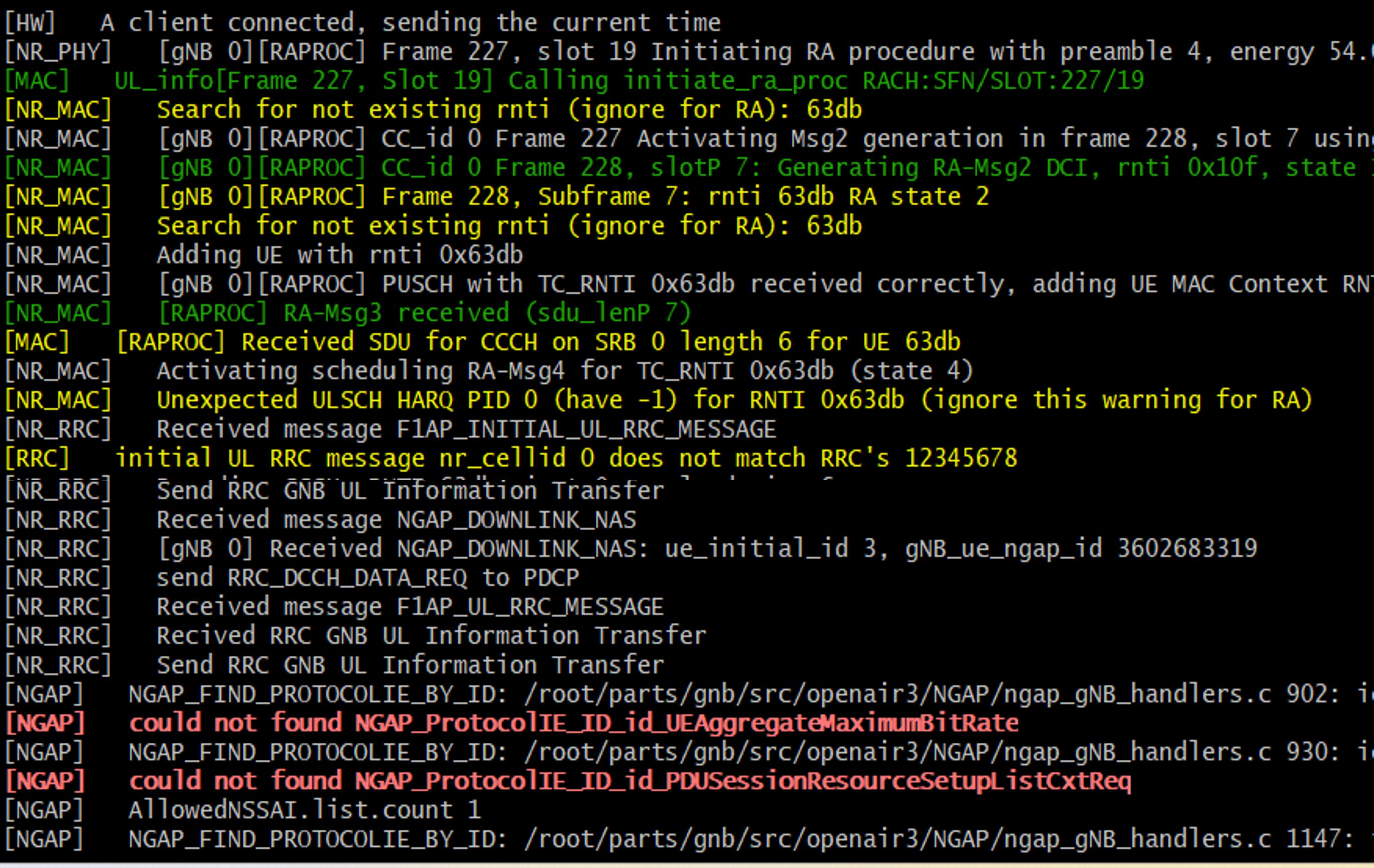}}
\caption{Gnode-B logs}
\label{fig:gnb_logs}
\end{figure}

As illustrated in Fig.~\ref{fig:gnb_logs}, the gNodeB log lines are color-coded. During filtering, only the colored lines (green, yellow, or red) are kept, as they represent the most relevant information.

\begin{figure}[h]
\centerline{\includegraphics[width=3.3in]{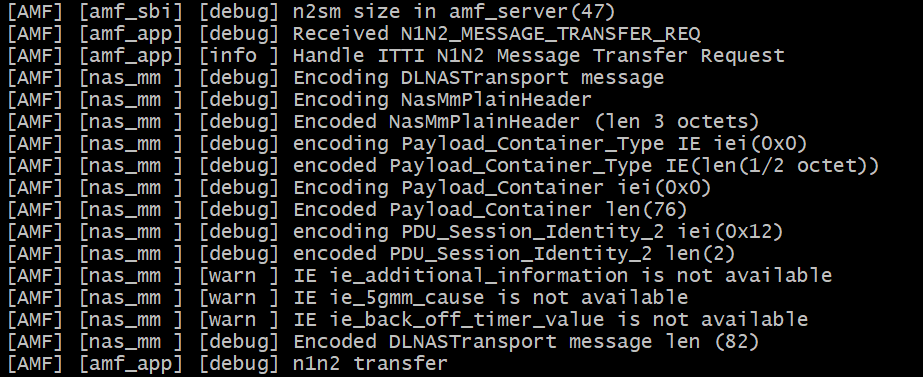}}
\caption{AMF logs}
\label{fig:amf_logs}
\end{figure}

Each log line of other network elements contains a tag indicating its level (see Fig.~\ref{fig:amf_logs}). The possible tags include \texttt{[info]} for informational messages, \texttt{[warn]} for warnings, \texttt{[debug]}, and \texttt{[trace]}. For LLM processing, only \texttt{[warn]} and \texttt{[info]} lines are retained.

\subsection{LLM Fine-Tuning}

As described in the data-gathering section, the following data were collected after injecting a fault, concatenated, and combined with a system prompt. The complete input was then provided to the LLM, which was instructed to identify the source of the problem:

\begin{itemize}
    \item \textbf{Pod status:} Output of \texttt{kubectl get pods}, listing \texttt{NAME}, \texttt{READY}, \texttt{STATUS}, \texttt{RESTARTS}, and \texttt{AGE}. Used to detect \textit{PodFailure} and \textit{PodKill} conditions.
    \item \textbf{Pod logs:} Logs from individual components (e.g., \texttt{amf}, \texttt{smf}, \texttt{upf}, \texttt{db}, \texttt{gnb}, \texttt{ue1}, \texttt{ue3}), such as \texttt{oai-amf\_filtered.txt} and \texttt{oai-db\_filtered.txt}.
    \item \textbf{Pod events:} Kubernetes event outputs.
    \item \textbf{RTT data:} Latency statistics for each UE before and after fault injection, e.g.,\\
    \texttt{rtt min/avg/max/mdev = 29.168/36.242/52.845/8.537 ms}.
\end{itemize}

Network elements include \texttt{amf}, \texttt{smf}, \texttt{upf}, \texttt{db} (database), \texttt{gnb}, and user equipments (UEs) such as \texttt{ue1} and \texttt{ue3}.

In the system prompt, each component of the data and the corresponding fault types were briefly described along with the logic for fault detection. A simplified version of the system prompt is shown in Fig.~\ref{fig:system_prompt}. The collected data, together with the system prompt, were provided to the LLM to infer and report network errors.

\begin{figure}[!t]
\centering
\begin{tcolorbox}[colback=gray!5,colframe=black,fonttitle=\bfseries,
  title=System Prompt (Fault Diagnosis Task Definition)]
The model acts as an expert \textbf{5G network fault analyzer} for \textbf{OpenAirInterface (OAI)} deployments on \textbf{Kubernetes},
identifying exactly one failure type among five: \emph{IOInjection, NetworkDelay, NetworkLoss, PodFailure,} or \emph{PodKill}. 
Input data include pod status (\texttt{kubectl get pods}), pod logs (AMF, SMF, UPF, DB, gNB, UEs), Kubernetes events, and RTT latency statistics.

\textbf{Detection logic:}
\begin{itemize}
  \item \textbf{IOInjection:} ``Unknown database 'oai\_5g'’’ in DB logs.  
  \item \textbf{NetworkDelay:} RTT after fault $\gg$ RTT before fault.  
  \item \textbf{NetworkLoss:} ``unknown RNTI’’ in gNB logs or missing RTT data.  
  \item \textbf{PodFailure:} pod status = \texttt{RunContainerError}, READY count decreases.  
  \item \textbf{PodKill:} pod status = \texttt{ContainerCreating} or \texttt{AGE < 2 min}.
\end{itemize}

Exactly one fault type occurs per analysis.  
The model outputs a concise diagnostic sentence only, e.g.,
\emph{``Yes, network loss, 60\% packets from RAN to AMF lost’’},  
without speculation or additional explanation.
\end{tcolorbox}
\caption{System prompt used for the 5G fault‑diagnosis model.}
\label{fig:system_prompt}
\end{figure}

\section{Experimental Results}

50\% of the data were used for training, 25\% for validation, and the remaining 25\% for testing.

\begin{figure}[h]
\centerline{\includegraphics[width=3.3in]{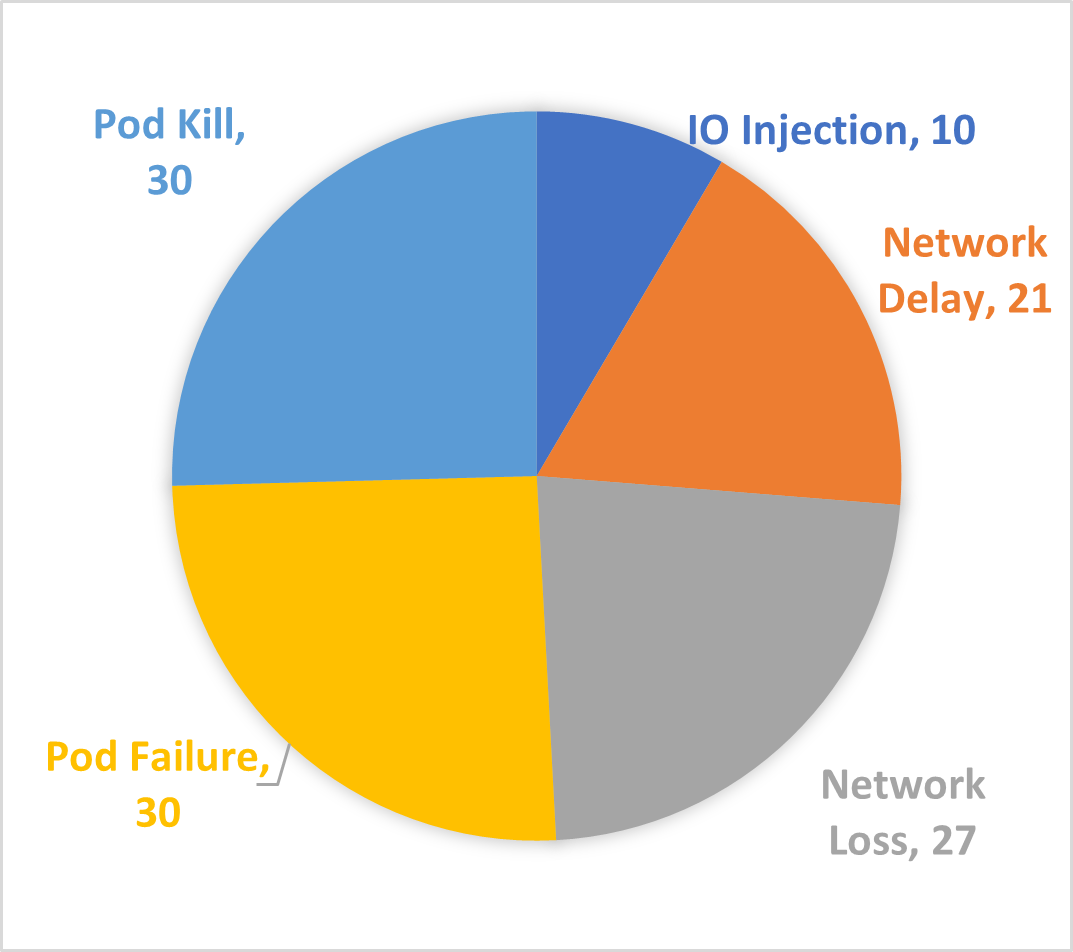}}
\caption{Distribution of fault-injection experiment types within the dataset}
\label{fig:fault_distribution}
\end{figure}

The GPT-4.1-Nano model was fine-tuned on the dataset using the OpenAI API. The dataset contains 118 experiments. As shown in Fig.~\ref{fig:fault_distribution}, the data volume for each fault type is uniformly distributed across all fault categories.

To evaluate the capability of the fine‑tuned LLM in network fault diagnosis, two evaluation modes were defined:

(i) Binary detection, determining whether any fault exists in the log response, and

(ii) Exact matching, verifying the entire predicted response against the labeled dataset to ensure correct identification of the specific fault type.

\subsection{Binary Fault Detection}
\begin{figure}[h]
\centerline{\includegraphics[width=3.3in]{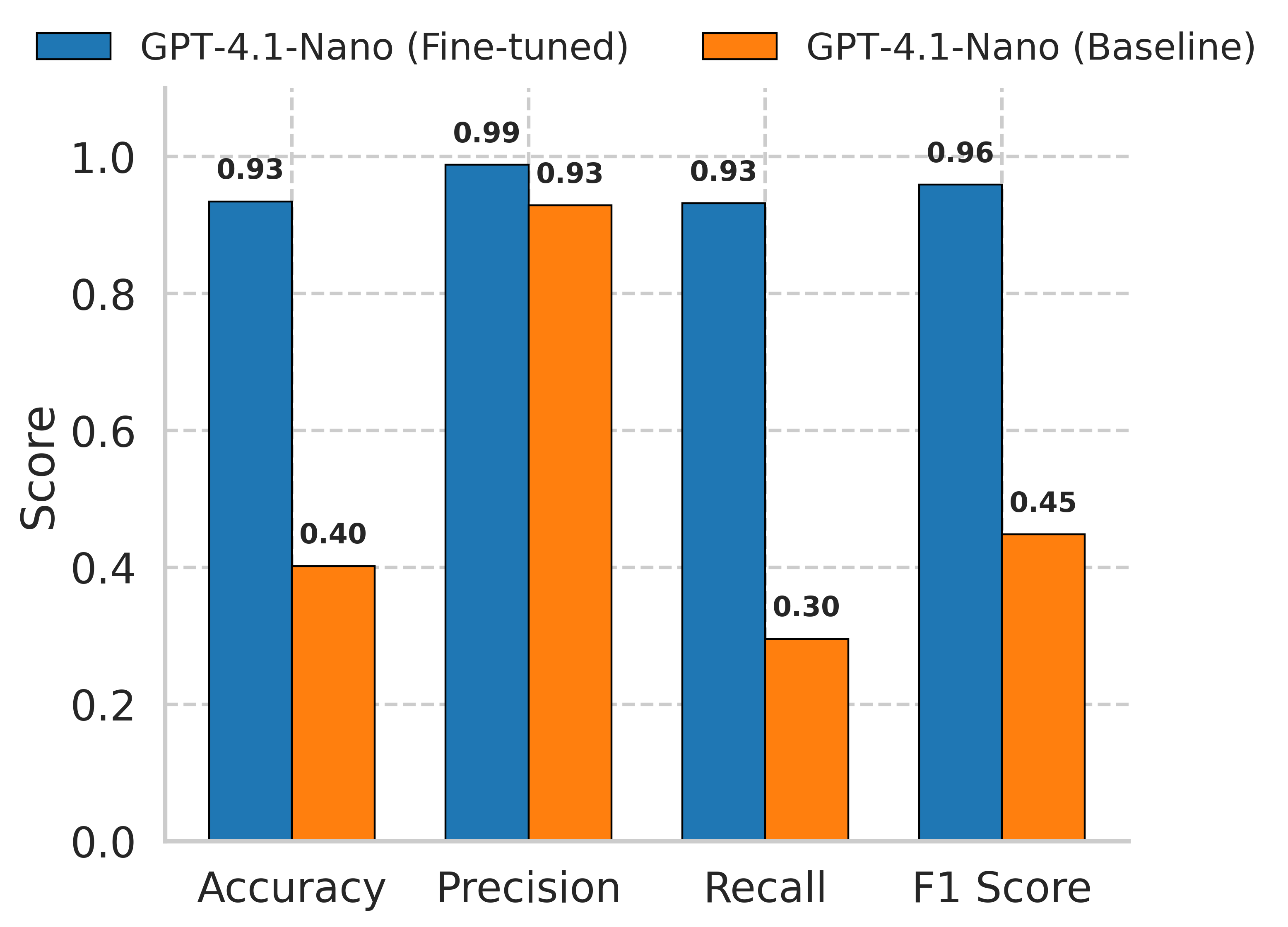}}
\caption{Binary comparison of fine tuned and base model in fault detection.}
\label{fig:result}
\end{figure}

Fig.~\ref{fig:result} summarizes overall performance for binary classification. The fine‑tuned model achieved 93\% accuracy and 95\% F1‑score, representing a substantial improvement over the non‑tuned baseline (40\% accuracy and 45\% F1‑score). This indicates that fine‑tuning significantly enhances the model’s consistency in detecting the presence of faults across diverse log inputs.

The improvement in recall from 30\% to 93\% indicates that the baseline model was weak in fault detection and often misclassified faulty networks as normal.

\subsection{Exact Fault Identification}
\begin{figure}[h]
\centerline{\includegraphics[width=3.3in]{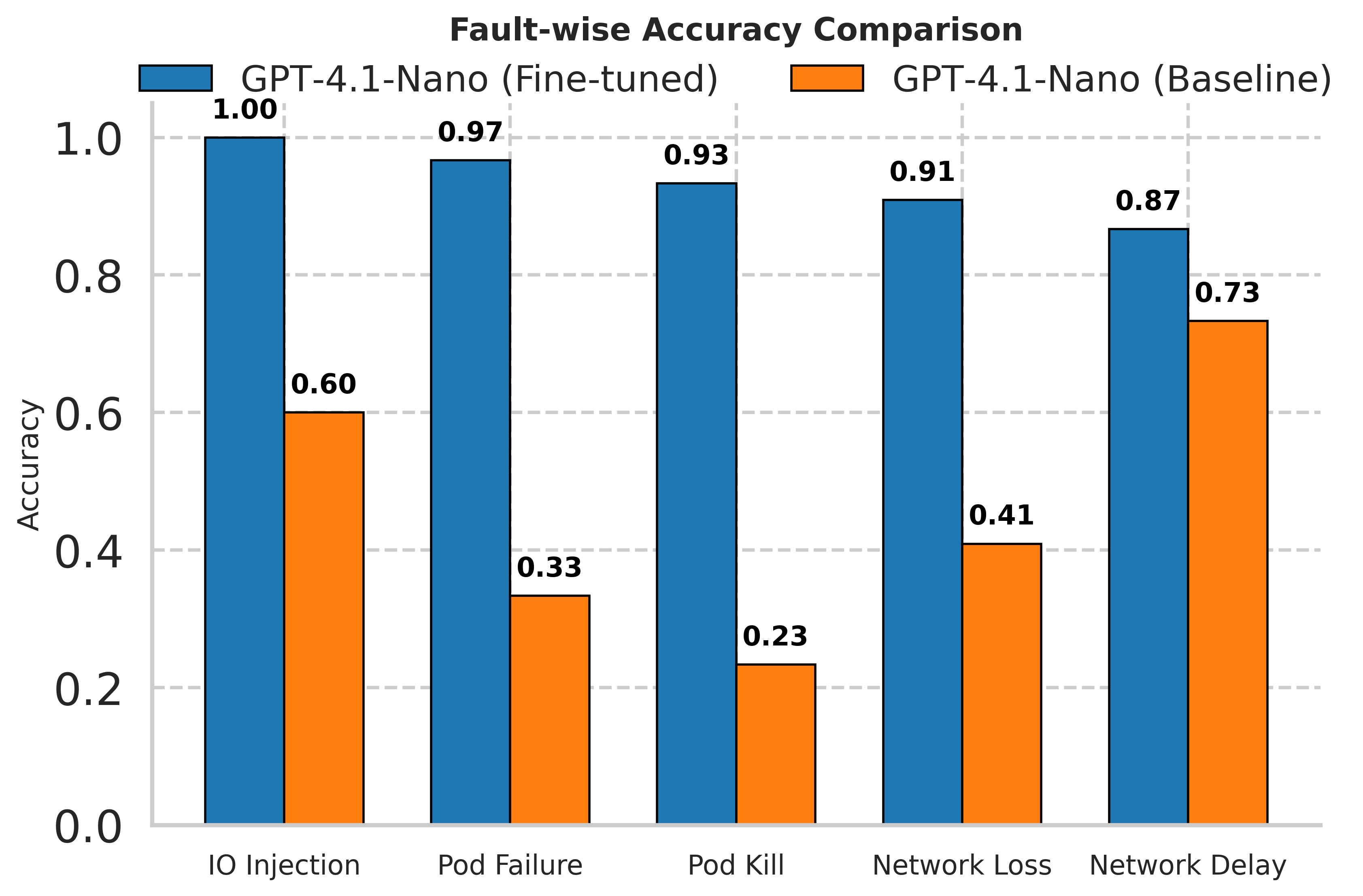}}
\caption{Binary comparison of fine tuned and base model in fault detection.}
\label{fig:faultwise_accuracy}
\end{figure}

Fig.~\ref{fig:faultwise_accuracy} illustrates the per-fault accuracy comparison between the fine-tuned and baseline \texttt{GPT-4.1-Nano} models across five fault categories. The fine-tuned model consistently outperforms its baseline counterpart for all fault types, achieving near-perfect accuracy for I/O~Injection (1.00) and Pod~Failure (0.97). 
The 100\% accuracy achieved for I/O~Injection is primarily attributable to the small sample size (10~instances) and the distinctive log pattern observed during this fault, where the database pod’s disk failed due to I/O~Injection and the unknown database \texttt{'oai'} was displayed in the DB logs. Consequently, the fine-tuned LLM identified this fault easily due to the clear and repetitive textual signature.
The improvement is particularly significant for Pod~Kill detection, where the baseline accuracy of 0.23 increases to 0.93 after fine-tuning, indicating a substantial enhancement in recognizing destructive pod-level terminations. Network-related faults, namely Network~Loss and Network~Delay, also exhibit strong performance stability with accuracies of 0.91 and 0.87, respectively, underscoring the model’s robustness to transient latency and packet-loss anomalies. These results confirm that the fine-tuning process effectively generalized across heterogeneous fault classes, yielding a more reliable and fault-sensitive LLM-based detection system.

\end{document}